\documentclass[aps,twocolumn,superscriptaddress]{revtex4}   

\newcommand{\dzz}{\ensuremath{d_{z^2}}}
\newcommand{\dxxyy}{\ensuremath{d_{x^2-y^2}}}
\newcommand{\dxy}{\ensuremath{d_{xy}}}

\newcommand{\dxz}{\ensuremath{d_{zx}}}
\newcommand{\dyz}{\ensuremath{d_{yz}}}

    \newcommand{\ThreeJSymbol}[6] {\ensuremath{
                 \left( \begin{array}{ccc}
                  {#1}&{#2}&{#3} \\
                  {#4}&{#5}&{#6}
                 \end{array} \right)
                                  }}

\usepackage{graphicx}
\usepackage{epsfig}
\usepackage[normalem]{ulem}
\usepackage{color}
\usepackage{bm}

\begin{document}

\title{Mixed configuration ground state in Iron(II) Phthalocyanine}
\author{Javier Fern\'{a}ndez-Rodr\'{\i}guez}

\affiliation{Department of Physics, Northern Illinois University, DeKalb, Illinois 60115, USA}
\affiliation{Advanced Photon Source, Argonne National Laboratory, 9700 South Cass Avenue, Argonne, Illinois 60439, USA}

\author{Brian Toby}

\affiliation{Advanced Photon Source, Argonne National Laboratory, 9700 South Cass Avenue, Argonne, Illinois 60439, USA}

\author{Michel van Veenendaal}

\affiliation{Department of Physics, Northern Illinois University, DeKalb, Illinois 60115, USA}
\affiliation{Advanced Photon Source, Argonne National Laboratory, 9700 South Cass Avenue, Argonne, Illinois 60439, USA}

\date{\today}

\begin{abstract}
We calculate the angular dependence of the x-ray linear and circular
dichroism  at the $L_{2,3}$ edges of $\alpha$-Fe(II) Phthalocyanine (FePc) thin films
using a ligand field model with full configuration interaction.
We find the best agreement with the experimental  spectra for a  mixed ground state of   $^3E_{g}(a_{1g}^2e_g^3b_{2g}^1)$ and $^3B_{2g}(a_{1g}^1e_g^4b_{2g}^1)$ with the two configurations coupled by the spin-orbit interaction.
The  $^3E_{g}(b)$ and $^3B_{2g}$ states have an easy axis and plane anisotropies, respectively.
Our model accounts for an easy-plane magnetic anisotropy and the  measured magnitudes of the in-plane orbital and spin moments.
The proximity in energy of the two  configurations
allows a switching of the magnetic anisotropy from easy plane
to easy axis with a small change in the crystal field, as recently observed
for FePc adsorbed on an oxidized Cu surface.
We also discuss the possibility of a  quintet ground state ($^5A_{1g}$ is 250~meV above the ground state) with planar anisotropy by manipulation of the Fe-C bond length by depositing the complex on a substrate that is subjected to a mechanical strain.
\end{abstract}

\maketitle

\section{Introduction}

Metal phthalocyanines (MPc's) have many technological
applications in catalysis~\cite{Leznoff1996}, photodynamic cancer therapy~\cite{Reddi1987}
and, given their semiconducting properties, in solar cells~\cite{PorphyrinHandbook}.
Another field of potential interest of MPc's is their use as magnetic materials~\cite{Miyoshi1973,Kirner1976},
with applications as magnetic storage devices, quantum computing,
and molecular spintronics~\cite{Bognani2008,Sanvito2007}.
Understanding the microscopic interactions that govern their magnetic properties
is of key importance in the design of functional materials.
In addition to the magnetic interactions (exchange) amongst the building blocks,
magnetic anisotropy has an important role in the magnetic properties of a material.
In planar MPc's the metal center is surrounded by the four pyrrolic nitrogens of the macrocycle in an
environment of $D_{4h}$ symmetry.
The $d$-orbitals of the metal-center are split into the four representations of the group~\cite{Ballhausen1962}:
$a_{1g} (\dzz)$, $b_{1g} (\dxxyy)$, $e_g$ ($\dxz$, $\dyz$), and $b_{2g} (\dxy)$.
Ligand-field models~\cite{Haverkort2012} can give the energy levels
of the individual orbitals and provide an adequate formalism for describing the
ground state and magnetic anisotropy of the individual
single molecule magnets ~\cite{Gatteschi2001}.

Fe(II)-Phthalocyanine (FePc) is a promising
candidate for its use as a magnetic material given its
strong magnetic anisotropy.  The tunability of the magnetization axis has been
the subject of recent study~\cite{Tsukahara2009,HuWu2013} for its
possible application in spintronics.
Despite having been a subject of study for several decades, 
the ground state configuration of FePc
is still  a matter of debate since it was originally proposed~\cite{Dale1968,Coppens1983}
as~$^3E_{g}(a_{1g}^1e_g^3b_{2g}^2)$ (in the following we abbreviate this ground state labeling as $^3E_{g}(a)$,
see footnote~\footnote{In $D_{4h}$ the labeling
$^3E_{g}$ can correspond to different ground states.
We denote the configurations $(a_{1g}^1e_g^3b_{2g}^2)$ and $(a_{1g}^2e_g^3b_{2g}^1)$
as $^3E_{g}(a)$ and $^3E_{g}(b)$ respectively.}).
Several density functional theory studies give different predictions.
Liao and Scheiner get a $^3A_{2g}(a_{1g}^2 e_g^2b_{2g}^2)$ ground state~\cite{LiaoScheiner2001}.
$^3A_{2g}$ has also been proposed based on x-ray measurements.~\cite{Kroll2012}
Marom {\it et al.} get $^3B_{2g}(a_{1g}^1e_g^4b_{2g}^1)$ or $^3A_{2g}$ depending on computational details
\cite{MaromKronik2009}.  Nakamura {\it et al.}~\cite{Nakamura2012} found $^3A_{2g}$ for isolated
FePc and  $^3E_g(a)$ for linear chains.
Recently, from multiplet calculations~\cite{Stepanow2011,Miedema2009} the ground state was
found to be $^3E_{g}(b)$.
Kuzmin~{\it et al.}~\cite{Kuzmin2013} also found $^3E_{g}(b)$ 
using a superposition crystal field model~\cite{Newman1989}.
The possibility of $^3E_{g}(b)$ and $^3B_{2g}$ lying very close
in energy and being mixed by spin-orbit coupling has been suggested~\cite{ReynoldsFiggis1991,Stepanow2011}.
A mixed quintet-triplet ground state has also been proposed~\cite{TholeLaanChem1988}.
M\"{o}ssbauer and x-ray dichroism measurements in thin films of $\alpha$-FePc~\cite{Filoti2006,Bartolome2010}
give valuable information about the electronic structure of FePc,
demonstrating that the complex has planar magnetic anisotropy, i.e., it is easier to magnetize
the molecule parallel to the plane and that the Fe ion has a large unquenched orbital moment
$m_L\approx 0.5\mu_B$.

In this paper, we use the multiplet model implemented in the xclaim code~\cite{xclaimPAPER,xclaimURL}
to calculate the $L_{2,3}$ edges x-ray spectra in FePc and determine the metal center ground state configuration
and crystal field energy levels.
By calculating the expectation values of the orbital and spin moments
we can determine the angular anisotropy and
estimate the errors affecting the application of the XMCD sum rules~\cite{Thole1992,Carra1993} in this system.
From a fit of the angular dependence of the x-ray absorption measurements in thin films of $\alpha$-FePc
by Bartolom\'{e} {\it et al.}~\cite{Bartolome2010} we determine the values of the
$D_{4h}$ crystal-field parameters and find a ground state of mixed
$^3E_{g}(b)$ and $^3B_{2g}$ character.
We discuss the magnetic anisotropies corresponding to the single configurations $^3E_{g}(b)$ and $^3B_{2g}$,
and propose $^3E_{g}(b)$ with easy axis anisotropy as the ground state in FePc adsorbed on an oxidized Cu surface.
From exact diagonalization we calculate the crystal-field excitations.  The presence of
a low-lying $^5A_{1g}$ configuration, makes it feasible to produce a quintet ground state with planar anisotropy by
manipulations of the Fe-C bond length.

\section{Ligand-field model}

For the Fe$^{2+}$ ion ($3d^6$) we use a ligand-field many-body hamiltonian with full
configuration-interaction taking into account
the Coulomb, spin-orbit coupling, crystal field and Zeeman interactions.
Hartree-Fock estimates of the radial part of matrix elements of the Coulomb interaction
in terms of Slater integrals $F^k$ and~$G^k$
and the spin-orbit coupling parameters $\zeta(3d)$
and $\zeta(2p)$ for the core $2p$ and valence $3d$ valence shells.
are obtained from Cowan's atomic multiplet program RCN~\cite{CowanBook,CowanURL}.
Their values are shown in table~\ref{TableParameters}.
The Slater integrals $F^k$ and~$G^k$ are reduced to 70\% of their Hartree-Fock values to
account for the effect of hybridization.

\begin{table}
\caption{\label{TableParameters} Hartree-Fock atomic parameters for the Fe$^{2+}$ ion in the base $3d^6$ and
excited $2p^53d^6$ atomic configurations in units of eV.  In the calculation of the spectra we apply a
70\% reduction to the Slater integrals.
}
\begin{center}
\begin{tabular}{lll}
parameter \hspace{0.6cm} & base config. & excited config. \\
\hline
$F^2(3d)$ &  10.966 &  11.779    \\
$F^4(3d)$ &  6.815  &  7.328     \\
$\zeta(3d)$ &  0.052  &  0.067     \\
$F^2(2p,3d)$  && 6.793  \\
$G^1(2p,3d)$ && 5.001  \\
$G^3(2p,3d)$ && 2.844  \\
$\zeta(2p)$    && 8.201  \\
\hline
\end{tabular}
\end{center}
\end{table}

The crystal field hamiltonian is written
in terms of Wybourne parameters~\cite{Mulak2000,Haverkort2005} as
\begin{equation}
\label{CFham}
   H_{\mathrm{CF}}= \sum_{k,q} B_{kq}C^{(k)}_q
\end{equation}
with $0\leq k\leq 2l$, $k$~an even integer and $-k\leq q \leq k$.  
$B_{kq}$  are the Wybourne parameters  and
$C^{(k)}_q$ are renormalized spherical harmonics $C^{(k)}_q = \sqrt{\frac{4\pi}{2k+1}}Y^k_q$.
The relationship $B_{k,-q}= (-1)^q B^{*}_{kq}$ holds because of the hermiticity of the hamiltonian.

For a $d$-shell in $D_{4h}$ symmetry, we can relate the $B_{20}$, $B_{40}$ and $B_{44}$
Wybourne parameters to the  Ballhausen notation 
$Dq$, $Ds$ and $Dt$~\cite{Ballhausen1962},
\begin{eqnarray}
  B_{20}&=&-7 Ds                             \nonumber\\
  B_{40}&=&21( Dq -Dt)                       \nonumber\\
  B_{44}&=&21\sqrt{\frac{5}{14}}Dq
\end{eqnarray}
In $D_{4h}$ the $d$-orbitals are split into four representations.
Their energies are, in terms of the Ballhausen parameters,
\begin{eqnarray}
\label{Edorbitals}
\epsilon_{a_{1g}}  &=&  6Dq-2Ds-6Dt   \nonumber \\
\epsilon_{b_{1g}}  &=&  6Dq+2Ds- Dt   \nonumber \\
\epsilon_{b_{2g}}  &=& -4Dq+2Ds- Dt   \nonumber \\
\epsilon_{e_{g }}  &=& -4Dq- Ds+4Dt   \nonumber
\end{eqnarray}

With the crystal field definition of Eq.~(\ref{CFham}) we can
calculate the effect of a rotation of the local ligand
environment by rotating the Wybourne parameters as a spherical tensor~\cite{Varsalovich}.  The crystal field
resulting from a rotation with Euler angles $\alpha\beta\gamma$ is expressed by
\begin{equation}
\label{generalBKQrotatated}
B'_{kq}= \sum_{-k\leq q'\leq k} B_{kq'}D^K_{q'q}(\alpha\beta\gamma) ,
\end{equation}
with the matrix elements $D^k_{q'q}(\alpha\beta\gamma)$ given by
\begin{equation}
D^k_{qq'}(\alpha\beta\gamma) = \exp(-iq\alpha) d^k_{qq'}(\beta)  \exp(-iq'\gamma) ,
\end{equation}
where $d^k_{qq'}(\beta)$ is a Wigner d-function.~\cite{Varsalovich}.
In this paper we consider a rotation of the sample 
about the $y$-axis and Equation~(\ref{generalBKQrotatated}) reduces to
\begin{equation}
B'_{kq}= \sum_{-k\leq q'\leq k} B_{kq'}d^k_{q'q}(\theta) ,
\end{equation}
where $\theta$ denotes the incidence angle of the x-rays with respect to the $c$-axis.
Particularizing for the case of a $d$-shell in $D_{4h}$ symmetry, the rotated crystal-field parameters can be written as
\begin{eqnarray}
B'_{2q}&=& B_{20}d^2_{0q}(\theta) \nonumber \\
B'_{4q}&=& B_{40}d^4_{0q}(\theta) + B_{44} [d^4_{4q}(\theta) + d^4_{-4q}(\theta) ]  .
\end{eqnarray}

\section{XAS and magnetic anisotropy}

\subsection{Fit to the experimental spectra}

We calculate the angular
dependence of the linearly polarized x-ray absorption and XMCD by rotating the crystal field
parameters and maintaining both the x-ray wavevector $\mathbf{k}$ and the 5~T applied magnetic field $\mathbf{H}$
parallel to the $z$-direction.
For the core-hole lorentzian broadenings we use the values
$\Gamma_L=0.2$ and $0.37$~eV for the Fe $L_{3}$ and $L_{2}$ edges~\cite{BookFuggle} and $\Gamma_{Gauss}=0.5$~eV
to account for the experimental resolution.
The additional broadening observed in the experimental spectra~\cite{Bartolome2010}
comes from the formation of electronic bands due to the columnar stacking of molecules in $\alpha$-FePc.

\begin{figure}
\begin{center}
\includegraphics[width=0.74\columnwidth,angle=90]{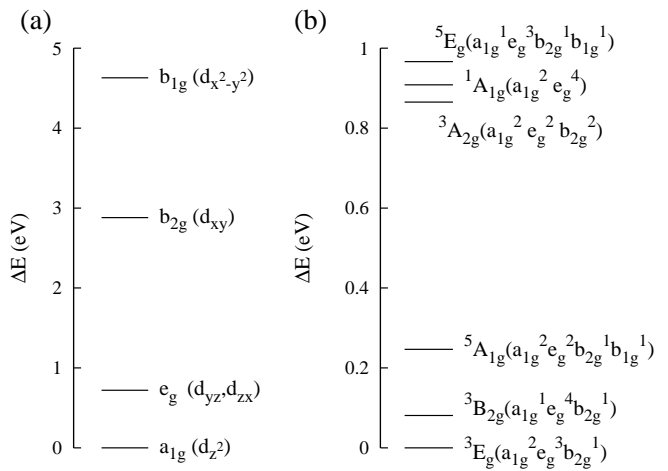}
\end{center}
\caption{\label{FigCF} (a) Crystal field levels of the d-orbitals in FePc
obtained from the fitting of the experimental spectra~\cite{Bartolome2010} and
(b) energy level diagram showing the ground state and lowest-lying $dd$ excitations
obtained from exact diagonalization of the many-body hamiltonian.
}
\end{figure}

\begin{figure}
\begin{center}
\includegraphics[width=0.999\columnwidth,angle=0]{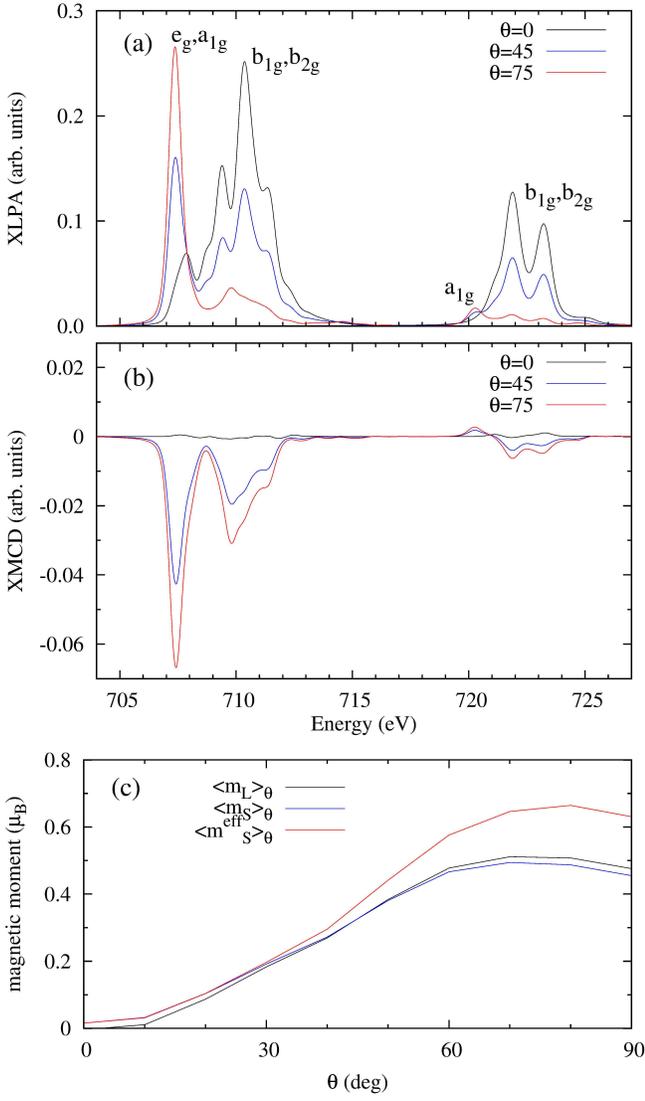}
\end{center}
\caption{\label{FIG_mix} Best fit to
the experimental x-ray linearly polarized absorption (XLPA) (a) and XMCD (b) at the Fe $L_{2,3}$ edges.
We show the calculated spectra for different x-ray incidence
angles $\theta=0, 45, 75^o$ with respect to the FePc $C_4$ axis.
The XLPA plot (a) includes the classification of the absorption peaks in terms of transitions to
valence orbitals belonging to different representations of $D_{4h}$.
The spectra correspond to a mixed $^3E_{g}(b)$ and  $^3B_{2g}$ ground state.
We also show in (c) the expectation values of the spin $\langle m_S\rangle_{\theta}$ and
orbital $\langle m_L\rangle_{\theta}$ magnetic moments along the direction of the applied magnetic field (H=5~T)
as a function of $\theta$.
We also include  the effective spin moment $\langle m^{\mathrm{eff}}_S\rangle_{\theta}$ that
results from applying the spin sum-rule to the calculated spectra.}
\end{figure}

The best fit to the experimentally measured spectra
corresponds to the crystal field parameters $Dq=0.175$, $Ds=0.970$, $Dt=0.150$~eV.
Figure~\ref{FigCF} shows the crystal field energy levels of the $d$-orbitals and the calculated
XAS spectra is plotted in Fig.~\ref{FIG_mix}.
We obtain a ground state of mixed  $^3E_{g}(e_g^3a_{1g}^2b_{2g}^1)$
and $^3B_{2g} (e_g^4a_{1g}^1b_{2g}^1)$ character.
The occupation of the orbitals in the mixed ground
state is $(a_{1g}^{1.7} b_{2g}^{1}e_g^{3.3})$, with part of the charge
of the $a_{1g}$ orbital being moved to the $e_{g}$ ($d_{zx}$, $d_{yz}$) orbitals.
When we switch off the spin-orbit interaction in our model, the mixture of the configurations
disappears and $^3E_{g}(b)$ is the ground state with $^3B_{2g}$ 80~meV higher in energy.
Fig.~\ref{FigCF}(b) shows a level diagram with the energies of the d-d excitations.
The next excitation is $^5A_{1g}(e_g^2a_{1g}^2b_{1g}^1b_{2g}^1)$ 250~meV above the ground state [Fig.~\ref{FigCF}(b)].
There are no other excitations within approximately 1~eV.
Our calculation gives a good account of the angular dependence of the lineraly polarized XAS and XMCD
when comparing it with the experimental measurements~\cite{Bartolome2010}.
The main shortcoming in our model is the absence of XMCD at $\theta=0$  (Fig.~\ref{FIG_mix}(b)).

By exactly diagonalizing the Hamiltonian we can get the 
zero field splitting (ZFS) of the mixed configuration ground state.
The excited states are 2.4, 3.6, 11, 47, 130 and 160 meV above the
ground state.  The states at 3.6 and 130 eV are doublets and the rest are singlets.
The applied magnetic field is not producing
a reordering of the zero-field energy levels, since their splittings are greater than the energy
changes induced by the magnetic field, in the order of ($\mu_BH \approx 0.3$~meV).

In addition to the x-ray spectra, we look at the magnetic anisotropy of the Fe ion.
In Fig.~\ref{FIG_mix}(c) we show the expectation values of the orbital $\langle m_L\rangle_{\theta}$
and spin $\langle m_S\rangle_{\theta}= 2\langle S\rangle_{\theta}$
magnetic moment components along the magnetic field as a function of $\theta$.
We calculate both the expectation value of the spin moment, and the spin effective 
$\langle m_S^{\mathrm{eff}}\rangle_{\theta}$, that is obtained by applying the spin sum rule~\cite{Carra1993}
to the calculated XMCD.  Several factors contribute to the discrepancy between 
the expectation value of the spin moment and its sum rule value:
the magnetic dipolar term $\langle T\rangle$~\cite{StohrKonig1995,Oguchi2004}, and the mixing of spectral weight between the $L_2$ and $L_3$ edges
occurring in early transition metals~\cite{Crocombette1996,Piamonteze2009}.
Another source of error is the fact that for practical applications
of the sum rule the isotropic intensity is approximated as the average of left and right
circularly polarized absorption~\cite{Chen1995} $I_{iso} =I_z+ I_++I_- \approx 3/2(I_++I_-)$.
An experimental measurement of the intensity with linear polarization along the $z$-axis $I_z$
would require the x-ray beam to be in the transverse direction.
At 5~T applied field, 
we get the moments in the $ab$-plane (see Fig.~\ref{FIG_mix}(c) at $\theta=90^o$) $\langle m^{ab}_L\rangle=0.5\mu_B$,
and $\langle m^{\mathrm{eff},ab}_{S}\rangle=0.7\mu_B$
in good agreement with the experimental values obtained by XMCD measurements~\cite{Bartolome2010}.

\subsection{Single configurations ground-states}

\begin{figure}
\begin{center}
\includegraphics[width=0.999\columnwidth,angle=0]{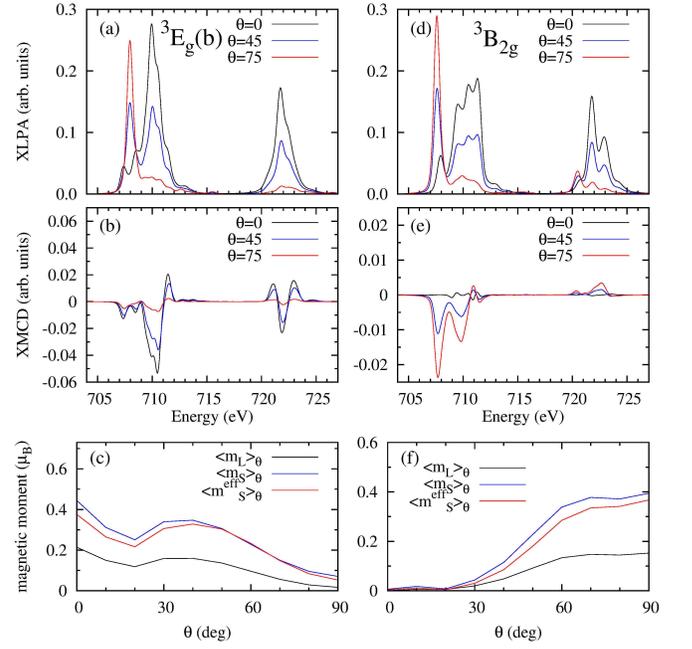}
\end{center}
\caption{\label{FIG_3E3B}  Dependence with the x-ray incidence angle $\theta$ of
the x-ray linearly polarized absorption (XLPA) and XMCD
for single-configuration $^3$E$_{g}(b)$ (a),(b) and $^3$B$_{2g}$ (d),(e) ground states.
(c),(f) Angular dependence of the expectation values of the spin $\langle m_S\rangle_{\theta}$ and
orbital $\langle m_L\rangle_{\theta}$ magnetic moments along the magnetic field
together with the spin magnetic moment determined from the XMCD sum rule $\langle m^{\mathrm{eff}}_S\rangle_{\theta}$. }
\end{figure}

In addition to the mixed configuration ground state that gives the best fit, we also show
the spectra corresponding to individual configuration ground states $^3E_{g}(b)$ and $^3B_{2g}$.
In our crystal field model, we can control the mixing in the ground state of
the two configurations by changing the energy positioning of the $d_{z^2}$
orbital with respect to the other $d$-orbitals (see
Eq.~\ref{Edorbitals}), and obtain single configuration ground states $^3E_{g}(b)$ or
$^3B_{2g}$.  In terms of Ballhausen parameters this corresponds to maintaining $Dq$ constant
and using the new parameters $Ds'=Ds-\Delta\epsilon_{z^2}/7$ $Dt'=Dt-\frac{3}{35}\Delta\epsilon_{z^2}$,
with $\Delta\epsilon_{z^2}$ the change in energy of the $d_{z^2}$ orbital.
For $^3E_{g}(b)$ we use $Ds=1.0275$, $Dt=0.1845$~eV [Fig.~\ref{FIG_3E3B}(a)-(c)] and for
$^3B_{2g}$, $Ds=0.913$, $Dt=0.116$~eV [Fig.~\ref{FIG_3E3B}(d)-(f)].

The two single configuration ground states would have different
magnetic anisotropies.  For $^3E_g(b)$ would be easy-axis with no magnetic moment in the plane [Fig.~\ref{FIG_3E3B}(c)]
and for $^3B_{2g}$ would be easy-plane [Fig.~\ref{FIG_3E3B}(f)].
A simple explanation for this behavior is given by the formalism for the magnetic anisotropy developed 
within a perturbative treatment of spin-orbit interaction.~\cite{Bruno1989,StohrKonig1995,vanderLaan1998}.  
Considering only spin-preserving excitations, the anisotropy energy
is proportional to the orbital moment, and we can discuss the anisotropy by looking at
the occupations of the single-particle orbitals in the ground state.~\cite{StohrKonig1995}.
For $^3B_{2g}$
$\dxy$ and $\dzz$ are singly occupied.  $\dzz$ cannot generate
orbital moment along the $c$-axis, and $\dxy$ can only generate
orbital moment along the $c$-axis from excitations to
$\dxxyy$, which is much higher in energy~(2~eV).
In $^3B_{2g}$  the orbital moment in the $ab$-plane comes from 
from $e_g\longrightarrow a_{1g}$ ($\dxz,\dyz\longrightarrow\dzz$) and $e_g\longrightarrow b_{2g}$
($\dxz,\dyz\longrightarrow\dxy$) excitations.
The case of $^3E_g(b)$ is different; since there is one hole in the $e_g$ orbitals with $m_l=\pm 1$,
orbital moment along the $c$-axis can be generated and the anisotropy is easy-axis.

\subsection{XAS angular dependence}

To understand the linearly polarized XAS spectral features in terms of transitions to valence orbitals we calculate
the angular dependence of the cross-section of dipolar transitions with the incidence angles to different
orbitals in the final state in terms of a single particle model.  
A similar discussion for the isotropic and XMCD spectra can be seen in Ref.~\onlinecite{KuzminHaynOison2009}.
For a representation $\Gamma$ of the point group $D_{4h}$ the absorption of linearly polarized x-rays is
\begin{equation}
I_{\Gamma}(\theta)=\sum_{m_j,\gamma} | \langle 3d,\gamma | D(\theta) | 2p,jm_j \rangle |^2
\end{equation}
%
where $\gamma$ label the $d$-shell orbitals belonging to the $\Gamma$ representation
and $m_j$ label the $p$-shell core states with $j=3/2$,~$1/2$
for the $L_3$,~$L_2$ edges.  $D(\theta)$ is the dipolar operator corresponding to
linear polarization in the $xz$-plane forming an angle $\theta$ with the $x$-axis:
\begin{equation}
D(\theta)=\cos\theta\frac{1}{\sqrt{2}}(D_{-1}-D_{+1}) +i\sin\theta D_{0}
\end{equation}
the component $D_q$~$(q=0,\pm 1)$ of the dipolar operator in spherical coordinates are
\begin{equation}
D_q=\sqrt{2}\sum_{m,m',\sigma}(-1)^m \ThreeJSymbol{2}{1}{1}{-m}{q}{m'}d^{\dagger}_{m,\sigma}p_{m',\sigma} .
\end{equation}

The resulting angular dependencies for linear polarization are given in
table~\ref{AngularDependenceXLPA}.  The relative 
intensities for the different representations are the same for the $L_3$ and $L_2$ edges.

In Fig.~\ref{FIG_mix}(a) we label the features of the XAS spectra according to transitions to valence orbitals
belonging to different representations of $D_{4h}$.
The sharp feature at the beginning of the $L_3$ edge (707 eV)
increases its intensity with $\theta$ and appears for both the
$^3B_{2g}$ and $^3E_g(b)$ ground states.  We assign it to transitions to the
$d_{z^2}$ or to the $e_g$ ($\dxz$, $\dyz$) orbitals contributing to the absorption at the same x-ray energy.
The transition at 720 eV at the low energy side of the $L_2$ edge visible in the linear polarization absorption
at $\theta=75^o$ can be
assigned to transitions to the $d_{z^2}$ orbital, given its appearance at high $\theta$
and the fact that it increases in intensity when the number of holes in $d_{z^2}$ increases.
For a pure $^3E_g(b)$ ground state [Fig.~\ref{FIG_3E3B}(a)], there are no holes in $d_{z^2}$ and the 720 eV peak
does not appear.
The tails of the $L_2$ and $L_3$ edges decrease at higher $\theta$.
We assign them to transitions to $b_{2g}(d_{xy})$ and $b_{1g}(d_{x^2-y^2})$ orbitals.
Both of them have the same angular dependence decreasing at higher incidence angle $\theta$.

\begin{table}
\caption{\label{AngularDependenceXLPA} Angular dependence of the linearly polarized x-ray absorption for single particle orbitals
belonging to the different representations $\Gamma$ of $D_{4h}$ as a function of the x-ray incidence angle $\theta$.}
\begin{center}
\begin{tabular}{ccl}
\hline
\hspace{0.3cm}$\Gamma$ \hspace{0.6cm}&\hspace{0.6cm} $d$-orbitals      \hspace{0.6cm} &\hspace{0.6cm} $I(\theta)$             \hspace{0.3cm}  \\
              \hline                                                                                                                               
\hspace{0.3cm}$a_{1g}$ \hspace{0.6cm}&\hspace{0.6cm} $d_{z^2}$         \hspace{0.6cm} &\hspace{0.6cm} $\frac{2}{45}(1+3\sin^2\theta)$\hspace{0.3cm} \\
\hspace{0.3cm}$b_{1g}$ \hspace{0.6cm}&\hspace{0.6cm} $d_{x^2-y^2}$     \hspace{0.6cm} &\hspace{0.6cm} $\frac{2}{15}\cos^2\theta$   \hspace{0.3cm} \\
\hspace{0.3cm}$b_{2g}$ \hspace{0.6cm}&\hspace{0.6cm} $d_{xy}$          \hspace{0.6cm} &\hspace{0.6cm} $\frac{2}{15}\cos^2\theta$   \hspace{0.3cm} \\
\hspace{0.3cm}$e_{g }$ \hspace{0.6cm}&\hspace{0.6cm} $d_{yz}$, $d_{zx}$\hspace{0.6cm} &\hspace{0.6cm} $\frac{1}{15}(1+\sin^2\theta)$ \hspace{0.3cm} \\
\hline
\end{tabular}
\end{center}
\end{table}

\section{Ground state changes and magnetic anisotropy switching}

Recently, the anisotropy of FePc has been reported to change from easy plane to easy axis when
adsorbed on an oxidized Cu$(110)$ surface~\cite{Tsukahara2009}.
Tsukahara {\it et al.}~\cite{Tsukahara2009} interpret the zero field splitting for isolated FePc in
terms of a simple model with an orbitally non-degenerate ground state
in which the zero-field splitting would only have two levels~\cite{Dale1968}
and attribute the observation of a more complex zero field splitting, and an easy-axis magnetic anisotropy of
FePc adsorbed on an oxidized Cu$(110)$ surface to the breaking of $D_{4h}$ symmetry.
However, a $^3E_g(b)$ configuration, with orbital degeration, can account for a complex zero field splitting
and an easy-axis magnetic anisotropy.

\begin{figure}
\begin{center}
\includegraphics[width=1.0\columnwidth,angle=0]{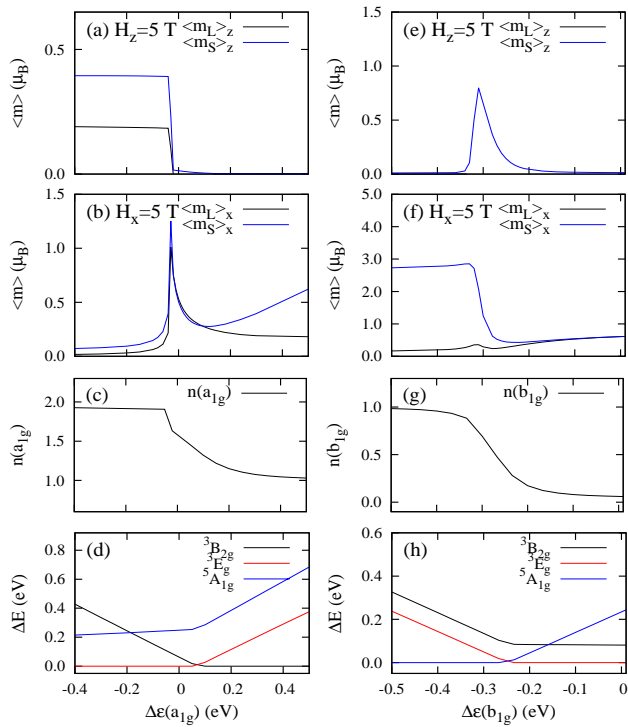}
\end{center}
\caption{\label{FIGevolGS} Changes in the ground state and magnetic anisotropy
as a function of the change in the crystal-field energies: (a-d) varying the energy
of the $a_{1g}$~$(\dzz)$ orbital and (e-h) changing the $b_{1g}$~$(\dxxyy)$ energy.
$\Delta\varepsilon(a_{1g})$ and $\Delta\varepsilon(b_{1g})$ correspond to the difference in
the single-particle orbital energies from the fitted values.
The plots show the orbital and spin magnetic moments generated for axial and in-plane applied fields,
the changes in the orbital occupations
and the energies relative to the ground state for the three configurations
$^3B_{2g}$, $^3E_g(b)$ and $^5A_{1g}$.  The splitting of the configurations
produced by spin-orbit coupling are not shown in (d) and (h).}
\end{figure}

In Fig.~\ref{FIGevolGS} we show the changes in the ground state and magnetic anisotropy
as a function of the perturbations of the crystal-field potential.  
The modifications of the $a_{1g}$ and $b_{1g}$ single-particle orbital energies
can be related to different physical effects: 
the $a_{1g}$ potential can be changed by axial ligand coordination~\cite{Wackerlin2012},
or when the complex is adsorbed in a surface~\cite{Tsukahara2009,HuWu2013}.  The $b_{1g}$ energy
would be modified by changes in the Fe-C bond length.
By changing $\varepsilon (a_{1g})$ [Fig.~\ref{FIGevolGS}(a-d)] the ground state changes between $^3E_g(b)$ and $^3B_{2g}$
and the anisotropy changes between easy-plane and easy-axis.  A reduction in $\varepsilon (a_{1g})$ as small as 0.1~eV
is enough to change the magnetic anisotropy to easy-axis.
The maximum of the in-plane generated moment corresponds to the region of mixed ground state,
and decreases when increasing $\varepsilon(a_{1g})$ [Fig.~\ref{FIGevolGS}(b)].
This can be easily understood, since the increase in $\varepsilon(a_{1g})$ diminishes the orbital moment
generated by $e_g\longrightarrow a_{1g}$ excitations.
It is worth noting that the mixed-configuration ground-state exists within a region of about~0.4~eV 
in the crystal field energies, where the $a_{1g}$ orbital has a non-integer
occupation [Fig.~\ref{FIGevolGS}(c)].

Changing the energy of the $b_{1g}$ orbital [Fig.~\ref{FIGevolGS}(e-h)] will produce a 
change in spin from triplet to a quintet~$^5A_{1g}$ with planar anisotropy.
Reducing $\varepsilon(b_{1g})$ by 0.2~eV will start populating the $b_{1g}$ orbital [Fig.~\ref{FIGevolGS}(g)].
Reductions beyond 0.3~eV produce a pure quintet ground state and saturate
the in-plane magnetic moment [Fig.~\ref{FIGevolGS}(f)].  Within a small intermediate region where
$^5A_{1g}$ is mixed with~$^3E_{g}$ magnetic moment can be generated
along the FePc axis [Fig.~\ref{FIGevolGS}(e)].
The changes in the energy of $b_{1g}$ can be related to changes in the Fe-C bond-length in FePc.
By considering the $a_{1g}$ crystal-field energy unaffected by the bond-length change,
we can consider the energy difference between $a_{1g}$ and $b_{1g}$ ($\approx 5$~eV) proportional
to $r^{-5}$~\cite{HarrisonBook}.
An increase of 1\% in the Fe-C bond-length ($\Delta r\approx 0.02$~\AA) would produce a quintet ground state.
This kind of spin transition produced by increasing in-plane bond-lengths
is feasible by depositing the complex on a substrate (graphene, polymers, etc.) and
exerting a mechanical strain on the substrate~\cite{Bhandary2011PRL}.

\section{Conclusions}

We have used a ligand field model with full configuration interaction to calculate the
magnetic properties and $L_{2,3}$ XAS spectra of FePc.
Our multiplet model gives a good account of the shape and angular
dependence of the experimental x-ray linearly polarized absorption and
XMCD spectra measured in thin films of $\alpha$-FePc.
The best fit to the experimental spectra corresponds to the
$D_{4h}$ crystal field parameters $Dq=0.175$, $Ds=0.970$, $Dt=0.150$~eV.
This corresponds a ground state of mixed $^3E_{g}(b)$ and $^3B_{2g}$ character,
as originally suggested by Reynolds {\it et al.}~\cite{ReynoldsFiggis1991}.
The two configurations are separated by a small amount of energy ($\approx 80$~meV) and the
spin-orbit  interaction produces a mixed ground state.
Although $^3E_{g}(b)$ (easy axis) is lower in energy, the mixing induced by the spin-orbit
produces a ground state with easy plane anisotropy.
The use of a full configuration interaction formalism makes possible to describe accurately
the magnetic properties of the system in the mixed configuration.

FePc is an excellent candidate for applications
in spintronics and information storage, with several ground states with
different magnetic properties accessible by
small changes in the ligand environment of the Fe ion.
The close proximity in energy of two configurations
with different easy magnetization axes makes it easy to manipulate the
magnetic anisotropy with very small changes in the crystal field.
The $^3E_g(b)$ configuration, with easy axis anisotropy, would be a plausible
ground state for FePc adsorbed on an oxidized Cu surface, where a change of the magnetization axis has
been reported.~\cite{Tsukahara2009}
In addition, the presence of a low-lying $^5A_{1g}$ configuration 250~meV above the
ground state, makes feasible to
produce a quintet ground state with planar anisotropy by expansions of the Fe-C bond length in the order of 0.02~\AA.
This can be achieved by depositing the complex in a substrate that is subjected to a mechanical strength.

The formalism used in this paper for the analysis of the XAS angular dependence can be applied to study
other systems and get information about the ground state and  $dd$ excitations.
The presence of low lying crystal-field excitations close to the ground state can
identify candidate systems for technological applications with tunable magnetic properties
where changes in the ligand environment would be able to change the ground state.

\section{Acknowledgments}

We thank F. Bartolom\'{e} for initially pointing us to the
XAS measurements in FePc.  We acknowledge useful discussions with
D. Haskel, U. Staub and J.A. Blanco.  This work was supported by the
U. S. Department of Energy (DOE), Office of Basic Energy Sciences,
Division of Materials Sciences and Engineering under Award
No. DE-FG02-03ER46097, the time-dependent x-ray spectroscopy
collaboration as part of the Computational Materials Science Network
(CMSCN) under grant DE-FG02-08ER46540, and NIU Institute for
Nanoscience, Engineering, and Technology. Work at Argonne National
Laboratory was supported by the U. S. DOE, Office of Science, Office
of Basic Energy Sciences, under contract No. DE-AC02-06CH11357.
This work utilized computational resources at NERSC, supported by the
U.S. DOE Contract No. DE-AC02-05CH11231.

\appendix
\section{Sum rule applicability for Fe$^{2+}$}

In this appendix we test the validity of the spin sum rule for
Fe$^{2+}$~($3d^6$).
The estimations of the sum rule error are useful 
for estimating the reliability of XMCD magnetization
measurements of Fe$^{2+}$ ions.  A particular case of interest is
SrFeO$_2$~\cite{Tsujimoto2007,Pruneda2008}, where the changes in
coordination leads to physical properties that are not well
understood.  We take several ground states as test cases and
apply the sum rule to calculated XMCD spectra and compare the
sum rule derived values with the ground state expectation of the spin.

We take into account triplet and quintet ground
states in tetragonal, octahedral and trigonal symmetries. 
It is worth noting that for Fe$^{(II)}$ in octahedral symmetry it is
not possible to obtain a triplet ground state~\cite{Miedema2009}.
Since the sum rule actually measures the expectation value of
the effective spin
$\langle SE_z \rangle = \langle S_z + \frac{7}{2}T_z \rangle$
which includes the magnetic dipole term $T_z$, we plot
the expectation values of both spin $S_z$ and spin effective $SE_z$ moments.

\begin{figure}
\begin{center}
\includegraphics[width=1.0\columnwidth,angle=0]{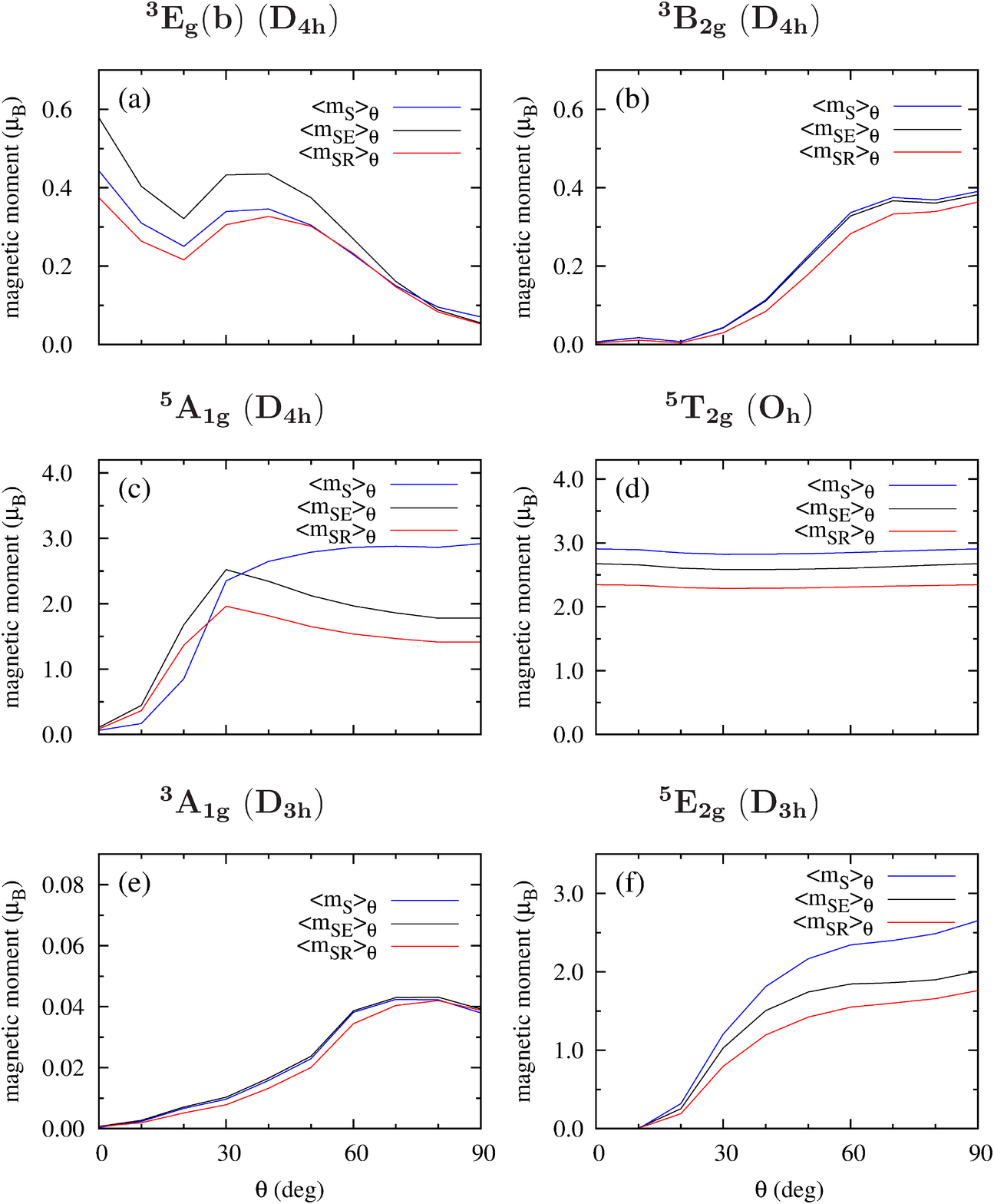}
\end{center}
\caption{\label{FIGsr} 
Spin magnetic moment $\langle m_S\rangle_{\theta}$ along the direction of the applied magnetic field (H=5~T)
as a function of the rotation angle  $\theta$ of the ligand environment for different ground states in 
tetragonal $(D_{4h})$, octahedral $(O_{h})$ and trigonal $(D_{3h})$ symmetries.
The spin moment values are plotted together with the effective spin
$\langle m_{SE}\rangle_{\theta}$ that includes the magnetic dipole contribution ($SE_z = S_z+\frac{7}{2}T_z$)
and the spin value $\langle m_{SR}\rangle_{\theta}$ determined from the
application of the spin sum rule to the calculated XMCD spectra.
}
\end{figure}

Fig.~\ref{FIGsr} shows the spin magnetic moment
$\langle m_S\rangle_{\theta}$ along the direction of the applied magnetic field (H=5~T)
for several ground states in $D_{4h}$, $O_{h}$
and $D_{3h}$ point groups.
The spin moment is shown together with the effective spin
$\langle m_{SE}\rangle_{\theta}$ that includes the magnetic dipole
contribution and the spin moment $\langle m_{SR}\rangle_{\theta}$
determined from applying the sum rule to the calculated XMCD spectra.
The values used for the crystal field parameters can be seen in the supplemental material addendum.


The two contributions to the sum rule error:
(the magnetic dipole term $T_z$ and the mixing of the $L_2$ and $L_3$ edges)
can be seen clearly in the plots.
The plots show a very good agreement of the sum rule with the
expectation value of the spin moment for triplet ground states with relative errors of less than 20\%
as can be seen in Fig.~\ref{FIGsr}(a), (b) and (e).
However for quintet ground states (Fig.~\ref{FIGsr} c,d and f) the relative error is much bigger and in the case of $^5A_{1g}$ in $D_{4h}$ symmetry
(Fig.~\ref{FIGsr}c) the spin moment is twice of the value measured by the sum rule.

\bibliography{ManuscriptFePc}
\end{document}